\documentclass[12pt]{article}
\usepackage{%
pstricks,%
pst-plot,%
epsf%
}                                  
\usepackage{cite}
\usepackage{array,dcolumn}      
\usepackage{axodraw}
\begin{document}
\begin{center}
{\large \bf   Bounds on masses of new gauge bosons in 
the 3 - 3 -1 models}
\footnote{Talk given at the {\it 4--th 
Rencontres du Vietnam~:  Physics at 
extreme energies}, Hanoi, 19-25 July 
2000.}\\
\vspace*{1cm}
     
{\bf $\mbox{Nguyen Anh Ky}^{a}$,  $\mbox{ Hoang  Ngoc Long}^{a}$,
$\mbox{Le Phuoc Trung}^{b}$, $\mbox{ Dang Van Soa}^{c}$} and 
{\bf $\mbox{Vo Thanh Van}^{a}$}\\
     
\vspace*{0.5cm}
     
{\it $^a$ Institute of
Physics, P. O. Box 429, Bo Ho, Hanoi 10000, Vietnam }\\ 
{\it $^b$ HCM  Branch, Institute of
Physics, Ho Chi Minh city, Vietnam }\\
{\it $c$ Department of Physics, Hanoi University of Mining and Geology}
     
\vspace*{1cm}
     
{\bf Abstract}\\[2mm]
\end{center}

{\small
\hspace*{0.5cm} Contribution from new gauge bosons in the 3 - 3 - 1
models to the anomalous magnetic moment of the muon, mass
difference of the kaon system and rare kaon decay are calculated
and numerically estimated. Bounds on masses of new gauge bosons:
bileptons and $Z'$ are derived.}

\section{Introduction}
\hspace*{0.5cm} The SuperKamiokande results \cite{suk} confirming 
non-zero neutrino mass call for the standard model (SM)  extension. 
Among the known extensions, the models based on the 
$\mbox{SU}(3)_C\times \mbox{SU}(3)_L \times \mbox{U}(1)_N$ gauge
group \cite{ppf,fhpp}(hereafter 3 - 3 - 1 models) have the following
intriguing features: firstly, the models are anomaly free only if
the number of families $N$ is a multiple of three. Further, from
the condition of QCD asymptotic freedom, which means $N < 5$, it
follows that $N$ is equal to 3. The second characteristic is that
the lagrangians of these models possess the Peccei-Quinn
symmetry naturally, hence the strong $CP$ problem can be solved in
an elegant way \cite{pal}. The third interesting feature is that
one of the quark families is treated differently from the other
two. This could lead to a natural explanation of the
unbalancing heavy top quarks in  the fermion mass
hierarchy. Recent analyses have indicated that signals
of new particles in this model, bileptons \cite{can} and exotic
quarks \cite{jm} may be observed at the Tevatron and the Large
Hadron Collider.\\

\hspace*{0.5cm} There are two main versions of 
the 3 - 3 - 1 models: the {\it
minimal} model in which all lepton components $(\nu, l, l^c)_L$ of
each family belong to one and same lepton triplet and a variant,
in which right-handed (RH) neutrinos  are
included, i.e. $(\nu, l, \nu^c)_L$ (hereafter we call it the model
with right-handed neutrinos \cite{rhnm,mpp}). New gauge bosons in the
minimal model are  bileptons
 ($Y^\pm, X^{\pm\pm}$) carrying lepton number $L =\pm 2$ and $Z'$.
In the second model, the bileptons with lepton number $L =\pm 2$
are singly--charged $Y^\pm$ and {\it neutral} gauge bosons $X^0,
X^{*0}$ , and both are responsible for lepton--number violating
interactions.  Thus, with the present group extension there are five 
new gauge bosons and all these particles are heavy. Getting mass 
limits for these particles is one of the central tasks 
of further studies.\\

\hspace*{0.5cm} In this report,  we summarize constraints on  new 
gauge boson masses using various experimental data, namely, 
the AMMM~\cite{kls}, mass difference of the kaon 
system~\cite{lv} and rare kaon decay~\cite{ltv}.\\

\section{Bounds on the bilepton masses from the AMMM}
\hspace*{0.5cm} The anomalous magnetic moments of the muon (AMMM) 
 is one of the most popular values in pursueing this
aim. Despite not competitive with the anomalous magnetic moment of
the electron (AMME) in precision, the AMMM is much more sensitive
to loop effects as well as ``New Physics" due to contributions
$\sim m_\mu^2$, i.e. $\sim (200)^2$ enhancement in the AMMM
relative to the AMME.
 Therefore the AMMM is a subject of both theoretical
and experimental investigations.

\subsection{The AMMM in the minimal model}
\hspace*{0.5cm} Before we go into the detailed calculation, 
let us recapitulate some basic elements of the model 
(for more details see \cite{dng}).
Three lepton components of each f amily  are in one triplet:
\begin{equation}
f^{a}_L =\left(
               \nu^a,\  l^a,\ (l^c)^a
                 \right)_L^T \sim (1, 3, 0),
\label{l} \end{equation} where $ a = 1, 2, 3$ 
is the family index. The
charged bileptons with lepton number $L = \pm 2$ are identified as
follows: $\sqrt{2}\ Y^-_\mu = W^4_\mu- iW^5_\mu , \sqrt{2}$\ $
X^{--}_\mu = W^6_\mu- iW^7_\mu $, and their couplings  to leptons
are given by \cite{framcal}
\begin{equation}
{\cal L}^{CC}_l = - \frac{g}{2\sqrt{2}}\left[ \bar{\nu}\gamma^\mu
(1- \gamma_5) C\bar{l}^{T}Y^-_\mu
 -  \bar{l}\gamma^\mu \gamma_5 C \bar{l}^T
X^{--}_\mu + \mbox{h.c.}\right]. 
\end{equation} 
It is to be noted that the
vector currents coupled to $X^{--}$, $X^{++}$ vanish due to Fermi
statistics. To get physical neutral gauge bosons one has to
diagonalize  their mass mixing matrix. That can be done in two
steps: At the first, the photon field $A_\mu$ and $Z,Z'$ are given
by \cite{dng} 
\begin{eqnarray} A_\mu  &=& s_W  W_{\mu}^3 + c_W\left(\sqrt{3}\
t_W  W^8_{\mu} +\sqrt{1- 3\ t^2_W}\ B_{\mu}\right),\nonumber\\
Z_\mu  &=& c_W  W_{\mu}^3 - s_W\left(\sqrt{3}\  t_W  W^8_{\mu}
+\sqrt{1- 3\ t^2_W}\ B_{\mu}\right),\nonumber\\ 
Z'_\mu & = &
\sqrt{3}\  t_W \  B_{\mu} - \sqrt{1- 3\ t^2_W}\ W^8_{\mu}.
\label{apstat} 
\end{eqnarray}
 In the second step, we get the physical
neutral gauge bosons $Z^1$ and $Z^2$ which are mixtures of  $Z$
and $ Z'$: 
\begin{eqnarray} Z^1  &=&Z\cos\phi - Z'\sin\phi,\nonumber\\ 
Z^2&=&Z\sin\phi + Z'\cos\phi. 
\end{eqnarray} 
The mixing angle $\phi$ is
constrained to be very small, therefore the $Z$ and the $Z'$ can
be safely  considered as the physical particles.

\hspace*{0.5cm} Now we calculate contributions 
from the bileptons and the $Z'$ to
the AMMM. It is known that heavy Higgs boson contribution to the
AMMM is negligible \cite{jw}, therefore the relevant diagrams are
depicted in Fig.1.
The first three diagrams come from the bileptons and their
contributions are found to be
\begin{equation}
\delta a_\mu^B = \frac{g^2 m_\mu^2}{24 \pi^2} \left(
\frac{16}{M_X^2} + \frac{5}{4 M_Y^2} \right), \label{dgbm} 
\end{equation}
where $M_X$, $M_Y$, $m_\mu$ stand for  masses of the doubly-,
singly-charged bileptons and of the muon, respectively. In the
limit $m_\mu << M_{Z'}$ where $M_{Z'}$ is the $Z'$ mass, the $Z'$
contribution has the form \cite{moore}
\begin{equation}
\delta a_\mu^{Z'} = \frac{m_\mu^2}{12 \pi^2 M_{Z'}^2}
 \left(g^{'2}_{V} - 5 g^{'2}_{A}\right).
\label{dgz} \end{equation} 
Following \cite{kls} (using Eq. (6) therein) 
we get coupling of the
muon to  the $Z'$
\begin{equation}
g'_{V}(\mu) =\frac{g}{c_W} \frac{3\sqrt{1 - 4
s_W^2}}{2\sqrt{3}},\ g'_{A}(\mu) = \frac{g}{c_W} \frac{\sqrt{1 - 4
s_W^2}}{2\sqrt{3}}. \label{hsm} \end{equation} 
Substituting (\ref{hsm}) into
(\ref{dgz}) we obtain the $Z'$ contribution
\begin{equation}
\delta a_\mu^{Z'} =\frac{g^2}{3 c_W^2}
 \frac{m_\mu^2}{12 \pi^2 M_{Z'}^2} ( 1-4 s_W^2 ).
\label{dgzm} 
\end{equation} 
Therefore the total contribution from  new gauge
bosons in the minimal version to the AMMM becomes~\cite{kls}
\begin{equation}
\delta a_\mu^{tm} = \frac{G_F m_W^2 m_\mu^2}{3\sqrt{2}\pi^2}
\left[ \frac{16}{ M_X^2} + \frac{5}{4 M_Y^2} +
 \frac{2 (1-4 s_W^2)}{3 c_W^2
M_{Z'}^2} \right], \label{dgzmt} 
\end{equation} 
where $G_F/\sqrt{2} = g^2/(8 m_W^2)$ is used.

\hspace*{0.5cm} Note that the $Z'$ gives a positive 
contribution to the AMMM,
while the $Z$ gives a negative one as it is well--known in the SM.
From Eq. (\ref{dgzmt}) it follows that the bilepton contributions
are dominant.
By the spontanous symmetry breaking (SSB) it follows
that ~\cite{kls} $|M_X^2 - M_Y^2| \leq 3 m_W^2$ . Therefore it is
acceptable to put $M_X \sim M_Y$ as it was done in \cite{fnsasaki}.
In this approximation,  Eq. (\ref{dgbm}) agrees with the original
result in \cite{fnsasaki}, and Eq. (\ref{dgzmt}) becomes
\begin{equation}
\delta a_\mu^{tm} = \frac{G_F m_W^2 m_\mu^2}{\sqrt{2}\pi^2} \left[
\frac{23}{4 M_Y^2} +
 \frac{2 (1-4 s_W^2)}{9 c_W^2
M_{Z'}^2} \right]. \label{dgzmtt} 
\end{equation}
A lower limit  $M_Y \sim 230$ GeV at  95\% CL  can be extracted by
the "wrong" muon decay $\mu \rightarrow e^- \nu_e \bar{\nu}_\mu $. 
Combining with the SSB, it follows \cite{dng} $M_{Z'} \geq 1.3$
TeV. With the quoted numbers ($ M_X = 180, M_Y = 230, M_{Z'} =
1300$ GeV), the contributions to $\delta a_\mu^{tm}$ from the
bileptons and the $Z'$ are $1.04 \times 10^{-8}$ and $7.76 \times
10^{-13}$, respectively. The bilepton contribution is in a range
of "New Physics" one \cite{cza}
 $\sim {\cal O}(10^{-8})$.

\hspace*{0.5cm} Putting a bound on "New Physics" contribution to the
AMMM~\cite{ki}
\begin{equation}
\delta a_\mu^{New\  Physics} = (7 \pm 8.6) \times 10^{-9},
\label{np} 
\end{equation} 
into the l.h.s of (\ref{dgzmtt}) we can obtain a
bound on $M_{Y}$. 
In \cite{kls} (see Fig. 2)  we plot $\delta a_\mu^{tm}$ as a
function of $M_Y$. For certainty we used $M_{Z'} = 1.3$ TeV 
quoted above. The horizontal lines are the upper and the lower  
limit from $\delta a_\mu^{New\  Physics}$.
 From the figure we get a lower mass limit on $M_{Y}$ to
be 167 GeV. We recall that this limit is in a range of  those
obtained from LEP data analysis ($M_Y \ge 120$ GeV) 
(see Ref. 21 of  \cite{kls}). 
In the near future, the E-821 
Collaboration at Brookhaven would reduce the experimental 
error on the AMMM to a few $\times
10^{-10}$. In Fig. 3 of \cite{kls} we see that 
$\delta a_\mu^{tm}$  cuts horizontal line I
($\sim 4 \times 10^{-10}$) and line II ($\sim 1 \times 10^{-10}$)
at $M_Y \approx 935$ GeV and
 $M_Y \approx 1870$ GeV, respectively. These lower bounds
are much higher than those from the muon experiments.

\subsection{The AMMM in the model with RH neutrinos} 
\hspace*{0.5cm} In this subsection we will calculate the
AMMM  in the model with RH neutrinos. Let us
recapitulate some basic elements of the model (for more details
see \cite{rhnm}). In this version the third member of the lepton
triplet is a RH neutrino instead of the antilepton $l^c_L$
\begin{equation}
f^{a}_L = \left(  \nu^a, l^a,  (\nu^c)^a \right)_L^T \sim (1, 3,
-1/3), l^a_R\sim (1, 1, -1). \label{lr} 
\end{equation} 
The complex new gauge
bosons $\sqrt{2}\ Y^-_\mu = W^6_\mu- iW^7_\mu ,\sqrt{2}\ X^0_\mu
=W^4_\mu- iW^5_\mu $ are responsible for lepton--number violating
interactions. Instead of the  doubly-charged bileptons
$X^{\pm\pm}$, here  we have  neutral ones $X^0$, $X^{0*}$. The SSB
gives the bilepton  mass splitting \cite{li}
\[ |M_Y^2 - M_X^2| \leq m_W^2.\]

As before one diagonalizes the mass mixing  matrix  of the neutral
gauge bosons by  two steps, and the last one is the same for both
versions. At the first step we have 
\begin{eqnarray} 
A_\mu  &=& s_W W_{\mu}^3
+ c_W\left(- \frac{t_W}{\sqrt{3}}\ W^8_{\mu}
+\sqrt{1-\frac{t^2_W}{3}}\  B_{\mu}\right),\nonumber\\
 Z_\mu  &=&
c_W  W^3_{\mu} - s_W\left( -\frac{t_W}{\sqrt{3}}\ W^8_{\mu} +
\sqrt{1-\frac{t_W^2}{3}}\  B_{\mu}\right), \nonumber \\
] Z'_\mu &=&
\sqrt{1-\frac{t_W^2}{3}}\  W^8_{\mu} +\frac{t_W}{\sqrt{3}} \
B_{\mu}. 
\label{apstatr} 
\end{eqnarray} 
Due to smallness of mixing angle
$\phi$ we can consider the $Z$ and  the  $Z'$ as the physical
particles. Due to its neutrality, the  bilepton  $X^0$ does not give a
contribution and in this case, the relevant diagrams are only two
last (c) and (d). The contribution from the singly-charged
bilepton and the $Z'$ in Fig. 1(c) and 1(d) is
\begin{equation}
\delta a_\mu^{tr} = \frac{G_F m_W^2 m_\mu^2}{12\sqrt{2}\pi^2}
\left\{ \frac{5}{M_Y^2} -  \frac{[5-(1-4 s_W^2)^2]}{2 c_W^2 (3 - 4
s_W^2)M_{Z'}^2} \right\}. \label{dgzrt}
 \end{equation}
In the considered
version the $Z'$ gives a negative contribution. However, the total
value in r.h.s of Eq. (\ref{dgzrt}) is positive (an opposite sign
happens when $M_{Z'} \leq 0.3\  M_Y$ which is excluded by the
SSB).

\hspace*{0.5cm} Putting the $Z'$ lower mass bound  
to be 1.3 TeV  \cite{dng} and
$M_Y = 230$ GeV we get the bilepton and the $Z'$ contributions to
$\delta a_\mu^{tr}$, respectively: $4.75 \times 10^{-10}$ and $-
7.87 \times 10^{-12}$. This implies that the contribution of the
new gauge bosons in the considered version is in two order smaller
than an allowed difference between theoretical calculation in the
SM and present experimental precision.
However, putting two previous values for $\delta a_\mu^{tr}$ we
get lower bounds on the bilepton masses to be about 250 GeV (I)
and 500 GeV (II) (see Fig. 4 in \cite{kls}).

\section{Constraint on $Z'$ mass
from the kaon mass difference $ \Delta \lowercase{m}_K$}
\hspace*{0.5cm} In the SM, flavour changing neutral current (FCNC) 
is completely suppressed by GIM mechanism  at tree level.
In the second or higher orders, this suppression is not complete
due to quark mass disparity \cite{InamiBuras}. In the $3-3-1$
model we can have FCNC even at tree level. In the left-handed
sector, since the third family has a different N charge from the
first and second family, their gauge couplings to $Z'$ are
different, leading to FCNC through the mismatch between weak and
mass eigenstates. Let us diagonalize mass matrices by three
biunitary transformations
\begin{eqnarray}
U'_L & = & V_L^U U_L,\   U'_R = V_R^U U_R,\nonumber\\
 D'_L & = &
V_L^D D_L,\   D'_R = V_R^D D_R, \label{tran}
\end{eqnarray}
where $U\equiv(u,c,t)^T$ and $D\equiv(d,s,b)^T$. The usual
Cabibbo-Kobayashi-Maskawa matrix is given by

\begin{equation}
V_{CKM} = V_L^{U+} V_L^D. \label{vckm}
\end{equation}
Using the unitarity of the $V^D$ and $V^U$ matrices, we get FCNC
interactions in down sector
\begin{equation}
{\cal L}^{NC}_{ds}=\frac{g c_W}{2 \sqrt{3-4 s_W^2}}
\left[V^{D*}_{Lid} V^D_{Lis}\right] \bar{d}_L \gamma^\mu s_L
Z'_\mu, \label{fcnc}
\end{equation}
where $i=(d,s,b)$;~  in our case $i=b$.

\hspace*{0.5cm} From the flavour-changing neutral current 
interaction (\ref{fcnc}),
we have the effective lagrangian
\begin{equation}
{\cal L}^{\Delta S=2}_{eff}=3D\frac{\sqrt{2} G_F c_W^4}{3-4 s_W^2}
\frac{m_Z^2}{m_Z'^2} \left[V^{D*}_{Lbd} V^D_{Lbs}\right]^2 \mid
\bar{d}_L \gamma^{\mu} s_L \mid^2. \label{effective}
\end{equation}
From the effective lagrangian, it is straightforward to get the
mass difference~\cite{lv}
\begin{equation}
\Delta m_K=\frac{4 G_F c_W^4}{3 \sqrt{2}(3-4 s_W^2)}
\frac{m_Z^2}{m_Z'^2} \left[V^{D*}_{Lbd} V^D_{Lbs}\right]^2 f_K^2
B_K m_K. \label{mass}
\end{equation}
It is expected that the $Z'$ contribution to $\Delta m_K$ is no
larger than observed values \cite{GaillardLee}. Using the
experimental values~\cite{pdg}
\begin{eqnarray}
\Delta m_K&=&3.489~^{+0.009}_{-0.009}\times 10^{-12} ~ \mbox{MeV},
\\ m_K&=&498 ~ \mbox{MeV},\\ \sqrt{B_K} f_K&=3D&135\pm19
~\mbox{MeV},
\end{eqnarray}
we have
\begin{equation}
m_{Z'}~\geq~2.63 \times 10^5 \eta_{Z'}[Re \mid V^{D*}_{Lbd}
V^D_{Lbs} \mid^2]^{1/2} \mbox{GeV},
\end{equation}
where $\eta_{Z'}=0.55$ is the leading order QCD
correction.
From the present experimental data we cannot impose constraints on
$V^{D*}_{Lbd}$ and $V^D_{Lbs}$. Using the Fritzch \cite{frit}
scheme
\begin{equation}
V^D_{ij} \approx \left( \frac{m_i}{m_j} \right)^{1/2},~~i<j,
\label{frit}
\end{equation}
where $i,~j$ are family indices, we get the bound on $Z'$ mass
\begin{equation}
m_{Z'}~ \geq ~1.02~ \mbox{TeV}. \label{zmfdm}
\end{equation}
\section{Constraint on $Z'$ mass from the rare decay
 $K^+\rightarrow \pi^+\nu\bar{\nu}$ }

\hspace*{0.5cm} In the SM the decay is loop-induced semileptonic FCNC
determined only by $Z^0$-penguin and box diagram. 
It is worthwhile to mention that the
photon-penguin contribution is absent in the decay since photon
does not couple to neutrinos.
We now move to discuss the semileptonic rare FCNC transition
$K^+\rightarrow \pi^+\nu\bar{\nu}$ in the framework of $3-3-1$
model and show how this decay can be used to get constraint on $
Z'$ mass.  The Feynman diagram contributing to the considered 
decay is depicted in Fig. 2.
In the 3 - 3 - 1  model, due to the FCNC 
interaction in (\ref{fcnc}) the decay can occur at 
tree level as in Fig. 2 of  \cite{ltv}. The decay
amplitude is given
\begin{eqnarray}
\nonumber {\cal M}_{\pi^+ \nu\bar{\nu}} &=& \frac{G_F}{2
\sqrt{2}}\frac{m_W^2}{m_{Z'}^2} V^{D*}_{Lbd}
 V^{D}_{Lbs}
\langle \pi^+(p_2)|\bar{s}_L \gamma_\mu d_L|K^+(p_1)\rangle 
\\ &\times & \bar{\nu}(k_1)\gamma^\mu(1 - \gamma_5) \nu(k_2),
\label{mt331}
\end{eqnarray}
where we have neglected $Z'$ momentum compared with its mass.
On the other hand, in the SM
the tree-level amplitude for the semileptonic
decay $K^+(p_1)\rightarrow \pi^0(p_2) e^+(k_1) \nu(k_2)$ is given
\begin{eqnarray}
\nonumber {\cal M}_{\pi^0 e^+ \nu}
 &=& \frac{G_F}{2 \sqrt{2}} V^{*}_{us}
\langle \pi^0(p_2)|\bar{s}_L \gamma_\mu u_L|K^+(p_1)\rangle \\ 
&\times & \bar{\nu}_e(k_1)\gamma^\mu(1 - \gamma_5) e(k_2).
\label{mtsm}
\end{eqnarray}
Isospin symmetry relates hadronic matrix elements in (\ref{mt331})
and (\ref{mtsm}) to a very good precision \cite{parsa}

\begin{equation}
\langle \pi^+(p_2)|\bar{s}_L \gamma_\mu d_L|K^+(p_1)\rangle =
\sqrt{2}\langle \pi^0(p_2)|\bar{s}_L \gamma_\mu
u_L|K^+(p_1)\rangle. \label{ht}
\end{equation}
Neglecting differences in the phase space of the two decays, due
to $m_{\pi^+}\neq m_{\pi^0}$ and $m_e \neq 0$, we obtain after
summation over three neutrino flavours~\cite{ltv}
\begin{equation}
\frac{Br(K^+ \rightarrow \pi^+\  \nu \  \bar{\nu})}{ Br(K^+
\rightarrow \pi^0\  e^+ \  \nu)} = 6 \left(
\frac{m_W^2}{m_{Z'}^2}\right)^2
\frac{|V_{Lbd}^{*D}V_{Lbs}^D|^2}{|V_{us}^*|^2}. \label{rat}
\end{equation}
Using the experimental data \cite{pdg}

\begin{eqnarray}
\label{me} \nonumber & &m_W =80.41~\mbox{GeV},~ |V_{us}| =
0.2196,~m_d=7~\mbox{MeV}, \\ & &  m_s=115~\mbox{MeV},~
m_b=4.3~\mbox{GeV}, \\ \nonumber & & Br(K^+ \rightarrow \pi^0\ e^+
\nu) = 4.42 \times 10^{-2}
\end{eqnarray}
and (\ref{frit}) we have
\begin{equation}
Br(K^+ \rightarrow \pi^+\nu \bar{\nu}) =
\frac{10923}{m_{Z'}^4\mbox{(GeV)}}. \label{massz}
\end{equation}
We notice that the standard model result after including
next-to-leading order QCD corrections for the decay is \cite{david}
\begin{equation}
0.79\times10^{-10}\leq Br(K^+ \rightarrow \pi^+\nu \bar{\nu})\leq
0.92 \times 10^{-10},
\end{equation}
while the present experimental values at Brookhaven \cite{brook}
\begin{equation}
Br(K^+ \rightarrow \pi^+\  \nu \ \bar{\nu})=1.5_{-1.2}^{+3.4}
\times 10^{-10}.\label{brook}
\end{equation}
Therefore, if $3-3-1$ symmetry is realized in nature, we can
expect that $Z'$ contribution to the decay $K^+ \rightarrow \pi^+
\nu \bar{\nu}$ is of order $10^{-10}$. Putting the central value
in (\ref{brook}) into (\ref{massz}), we get $m_{Z'}\simeq 2.3$
TeV. \\

\hspace*{0.5cm} Now let us consider the decay in the minimal model. 
In this model
the FCNC interaction is described by \cite{dng,dumm}
\begin{equation}
{\cal L}^{NC}_{ds}=\frac{g c_W}{2 \sqrt{3(1-4 s_W^2)}}
\left[V^{D*}_{Lbd} V^D_{Lbs}\right] \bar{d}_L \gamma^\mu s_L
Z'_\mu.
\end{equation}
Following the same steps as we have done in the r.h.n. model we
obtain
\begin{equation}
\frac{Br_m (K^+ \rightarrow \pi^+\  \nu \  \bar{\nu})}{ Br(K^+
\rightarrow \pi^0\  e^+ \  \nu)} = \frac{2}{3} \left(
\frac{m_W^2}{m_{Z'}^2}\right)^2
\frac{|V_{Lbd}^{*D}V_{Lbs}^D|^2}{|V_{us}^*|^2}. \label{ratmin}
\end{equation}
The index {\it m} indicates that the branching ratio is calculated
in the minimal model. Using (\ref{me}) we find

\begin{equation}
Br_m(K^+ \rightarrow \pi^+\  \nu \ \bar{\nu}) =
\frac{1212}{m_{Z'}^4\mbox{(GeV)}}.
\end{equation}
Using the  measured decay branching ratio in (\ref{brook}) we get
$m_{Z'}\simeq1.7$ TeV. This result is consistent with constraints
in \cite{dng} which come from muon decay and neutrino-nucleus
scattering. It is worthwhile mentioning that the branching ratio
is not sensitive to the value of $\sin^2\theta_W$, while the
expression of $\Delta m_K$ is very sensitive to $\sin^2\theta_W$.
\\

\hspace*{0.5cm}   In conclusion we emphasize that 
the new gauge bosons in the 3 - 3 - 1
models have lower mass limit in the range of TeV scale, and these 
models can be checked at the near future experiments.

\newpage
\begin{center}
\begin{picture}(260,50)(-5,0)
\Photon(1,20)(1,45){2}{3}
\ArrowLine(1,20)(-20,-10)
\ArrowLine(22,-10)(1,20)
\Photon(-20,-10)(22,-10){2}{5}
\ArrowLine(22,-10)(43,-40)
\ArrowLine(-41,-40)(-20,-10)
\Text(1,-45)[]{(a)}
\Text(10,30)[]{$\gamma$}
\Text(-18,10)[]{$\mu^-$}
\Text(-38,-20)[]{$\mu^-$}
\Text(26,10)[]{$\mu^-$}
\Text(46,-20)[]{$\mu^-$}
\Text(0,-20)[]{$X^{--}$}
\Photon(185,20)(185,45){2}{3}
\Photon(185,20)(168,-10){2}{4}
\Photon(185,20)(202,-10){2}{4}
\ArrowLine(202,-10)(168,-10)
\ArrowLine(151,-40)(168,-10)
\ArrowLine(202,-10)(219,-40)
\Text(185,-45)[]{(b)}
\Text(195,30)[]{$\gamma$}
\Text(165,10)[]{$X^{--}$}
\Text(151,-20)[]{$\mu^-$}
\Text(212,10)[]{$X^{- -}$}
\Text(227,-20)[]{$\mu^-$}
\Text(185,-18)[]{$\mu^{-}$}
\Photon(1,-80)(1,-105){2}{3}
\Photon(1,-105)(-19,-140){2}{4}
\Photon(1,-105)(22,-140){2}{4}
\ArrowLine(22,-140)(-19,-140)
\ArrowLine(22,-140)(43,-170)
\ArrowLine(-40,-170)(-19,-140)
\Text(1,-175)[]{(c)}
\Text(10,-100)[]{$\gamma$}
\Text(-19,-120)[]{$Y^-$}
\Text(-34,-150)[]{$\mu^-$}
\Text(27,-120)[]{$Y^-$}
\Text(45,-150)[]{$\mu^-$}
\Text(1,-149)[]{$\nu_\mu$}
\Photon(185,-80)(185,-105){2}{3}
\ArrowLine(164,-140)(185,-105)
\ArrowLine(185,-105)(206,-140)
\Photon(164,-140)(206,-140){2}{5}
\ArrowLine(145,-170)(164,-140)
\ArrowLine(206,-140)(224,-170)
\Text(185,-175)[]{(d)}
\Text(195,-100)[]{$\gamma$}
\Text(165,-120)[]{$\mu^-$}
\Text(150,-150)[]{$\mu^-$}
\Text(210,-120)[]{$\mu^-$}
\Text(227,-150)[]{$\mu^-$}
\Text(185,-150)[]{$Z'$}
\Text(87,-210)[]{ Figure  1: 
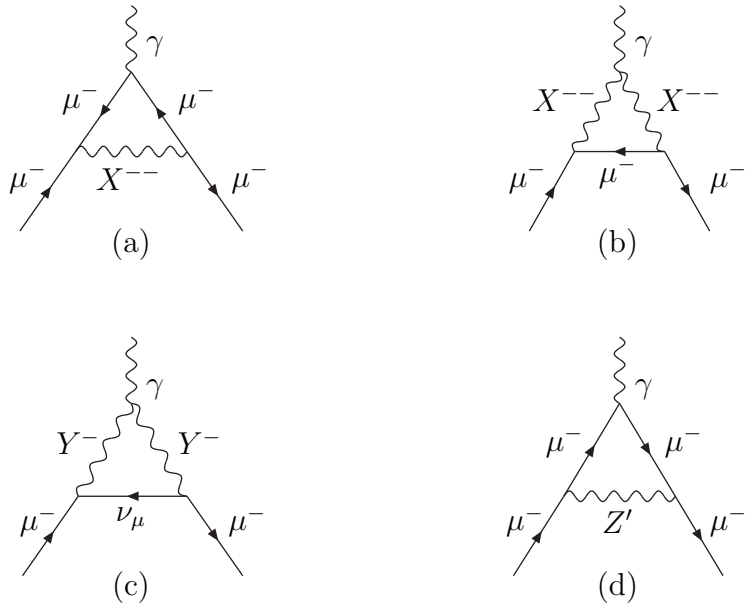
Diagrams contributing to the $(g_{\mu}-2)/2$.}
\end{picture}
\end{center}
\vspace*{10cm}

\begin{center}
\begin{picture}(260,50)(-5,0)
\ArrowLine(20,50)(31,10)
\ArrowLine(31,10)(20,-30)
\Photon(31,10)(110,10){2}{6}
\ArrowLine(110,10)(130,50)
\ArrowLine(130,-30)(110,10)
\ArrowLine(10,50)(10,-30)
\Text(75,18)[]{$ Z' $}
\Text(14,58)[]{$K^+$}
\Text(14,-36)[]{$\pi^+$}
\Text(142,55)[]{$\nu$}
\Text(142,-35)[]{$\bar{\nu}$}
\Text(4,30)[]{$u$}
\Text(4,-20)[]{$u$}
\Text(29,-20)[]{$\bar{d}$}
\Text(30,30)[]{$\bar{s}$}
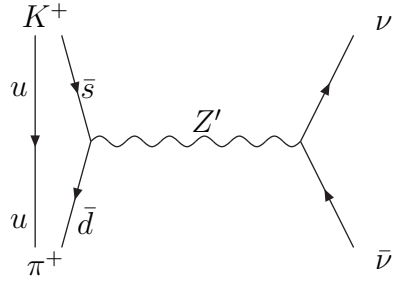
\Text(90,-80)[]{ Figure 2: Feynman diagram for 
$K^+ \rightarrow  \pi^+ \nu \bar{\nu}  $} 
\Text(90,-90)[]{ in the  3 - 3 - 1 models. } 
\end{picture}
\end{center}
\vspace*{4cm}
     
\end{document}